\documentclass[12pt]{article}
\usepackage{amssymb,amsmath}
\usepackage[dvipdfmx]{graphicx}
\usepackage{graphicx}
\usepackage{subfigure}
\usepackage{ytableau}
\usepackage{comment}
\usepackage[vcentermath]{youngtab}
\usepackage{hyperref}

\DeclareMathOperator*{\Tr}{{\rm Tr}}

\DeclareMathOperator{\Li}{{\rm Li}}

\begin{document}

\thispagestyle{empty}
\begin{flushright}
RUP-23-25
\end{flushright}
\vskip1cm
\begin{center}
{\bf {\LARGE Excitations of bubbling geometries \\
\vskip0.75cm 
for line defects}}

\vskip1.5cm

Yasuyuki Hatsuda\footnote{yhatsuda@rikkyo.ac.jp}

\bigskip
{\it Department of Physics, Rikkyo University, \\
Toshima, Tokyo 171-8501, Japan
}

\bigskip
\bigskip

Tadashi Okazaki\footnote{tokazaki@seu.edu.cn}

\bigskip
{\it Shing-Tung Yau Center of Southeast University,\\
Yifu Architecture Building, No.2 Sipailou, Xuanwu district, \\
Nanjing, Jiangsu, 210096, China
}

\end{center}

\vskip1.25cm
\begin{abstract}
The half-BPS Wilson line operators in the irreducible representations labeled 
by the Young diagrams for $\mathcal{N}=4$ $U(N)$ super Yang-Mills theory have gravity dual descriptions. 
When the number $k$ of boxes of the diagram grows as $k\sim N^2$, the bubbling geometries emerge.
We evaluate the spectra of quantum fluctuations on the bubbling geometries from the large $N$ and large $k$ limit of the supersymmetric indices decorated by the Wilson lines. 
The spectra of excitations over multi-particle $1/8$- and $1/2$-BPS states agree with the numbers of conjugacy classes of general linear group over finite fields 
while degeneracies of single particle BPS states are given by the general necklace polynomial. 
The bubbling geometry exhibits a new class of asymptotic degeneracy of states. 
\end{abstract}


\section{Introduction}
In the AdS/CFT correspondence \cite{Maldacena:1997re}, 
the half-BPS Wilson line operators in $\mathcal{N}=4$ super Yang-Mills (SYM) theory with gauge group $U(N)$ is conjectured to 
have the dual gravitational description in terms of Type IIB branes and strings in $AdS_5\times S^5$. 
The irreducible representations (irreps) of $U(N)$ in which Wilson line operators transform are labeled by the Young diagrams. 
The Wilson line in the fundamental representation is dual to the Type IIB fundamental string wrapping $AdS_2$ in the global $AdS_5$ \cite{Maldacena:1998im,Rey:1998ik}. 
For rank-$k$ symmetric and antisymmetric representations, 
the Wilson lines correspond to $k$ fundamental strings ending on an extra D3-brane wrapping $AdS_2\times S^2$ $\subset$ $AdS_5\times S^5$ 
\cite{Drukker:2005kx} 
and those attached to a D5-brane wrapping $AdS_2\times S^4$ $\subset$ $AdS_5\times S^5$ 
\cite{Yamaguchi:2006tq} respectively. 
For more general representations described by the Young diagram with $k$ boxes, 
the gravity dual configurations are realized by $k$ fundamental strings terminating on multiple D3- and D5-branes \cite{Gomis:2006sb}. 

In the large representations such that the Young diagrams have a large number of boxes, one encounters attractive geometries. 
When $k$ is large while $k/N$ is fixed, the gravity dual geometries are microscopically described by probe branes with fluxes 
whose backreaction on the supergravity solutions is neglectable. 
Beyond that, when $k$ is large while $k/N^2$ is kept, they lead to new geometries as more general supergravity solutions, the bubbling geometries. 
They are constructed as $AdS_2\times S^2\times S^4$ fibrations over 
a two-dimensional Riemann surface $\Sigma$ with boundary $\partial \Sigma$ which can develop multiple bubbles of cycles carrying fluxes 
in such a way that the fiber becomes singular and either $S^2$ or $S^4$ shrinks at the boundary of the surface. 
This is the Wilson line version of the bubbling geometry \cite{Yamaguchi:2006te,Lunin:2006xr,Okuda:2007kh,DHoker:2007mci,Okuda:2008px,Gomis:2008qa,Benichou:2011aa,Fiol:2013hna,Aguilera-Damia:2017znn} 
which generalizes the Lin-Lunin-Maldacena bubbling geometry \cite{Lin:2004nb} for the half-BPS local operators. 

The spectrum of the quadratic fluctuations of the fundamental string wrapping $AdS_2$ in $AdS_5\times S^5$ 
were derived using the Green-Schwarz formalism \cite{Drukker:2000ep}. 
Also the spectrum of quantum excitations of the probe brane descriptions, a D3-brane wrapping $AdS_2\times S^2$ and a D5-brane wrapping $AdS_2\times S^4$ 
are addressed from the action for the probe D-brane with fluxes \cite{Faraggi:2011ge,Faraggi:2011bb}. 
However, for the bubbling geometries, 
the probe brane approximation is no longer valid due to a fully backreacted supergravity background and little is known about excitations. 

In this Letter we present excitations on the bubbling geometries  
by analyzing the Schur line defect correlators for the dual $\mathcal{N}=4$ SYM theory 
\cite{Gang:2012yr,Drukker:2015spa,Gaiotto:2020vqj,Hatsuda:2023iwi,Guo:2023mkn,Hatsuda:2023imp}, 
which decorate the Schur index \cite{Gadde:2011ik,Gadde:2011uv}. 
Our method is surprisingly powerful to evaluate the gravity indices and our results reveal remarkable relationship 
between the unknown quantum fluctuations of the bubbling geometries and the combinatorial objects. 

\section{Gravity dual of the Wilson lines}
The half-BPS Wilson line in $\mathcal{N}=4$ SYM theory breaks the four-dimensional conformal symmetry $SU(2,2)$ 
down to $SU(1,1)$ $\times$ $SU(2)$ and the R-symmetry $SO(6)$ down to $SO(5)$ so that it can preserve the 
one-dimensional superconformal symmetry $OSp(4^*|4)$. 
$\mathcal{N}=4$ $U(N)$ SYM theory is realized as the low-energy effective theory of $N$ D3-branes in Type IIB string theory. 
The Wilson line in the fundamental representation is described by a fundamental string \cite{Maldacena:1998im,Rey:1998ik}. 
Unlike a proper open string, it also obeys Dirichlet boundary condition along the direction parallel to the D3-banes \cite{Drukker:1999zq}. 
The Wilson line in the $k$-th antisymmetric representation corresponds to $k$ fundamental strings between 
the stack of D3-branes and a single D5-brane \cite{Yamaguchi:2006tq}. 
Each string must terminate on a distinct D3-brane due to the s-rule \cite{Hanany:1996ie}. 
This explains why the number $k$ should be at most $N$. 
For the $k$-th symmetric representation, 
the Wilson line is realized by introducing an extra D3-brane parallel to the stack of D3-branes between which $k$ fundamental strings stretch \cite{Drukker:2005kx}. 
More generally, 
the irrep of $U(N)$ is labeled by the Young diagram 
for which there are two gravity dual descriptions of the Wilson line in terms of fundamental strings 
which terminate either on D3- or D5-branes \cite{Gomis:2006sb}. 
The D3-brane description is obtained by associating $i$-th row of the diagram with $k_i$ boxes 
to $i$-th D3-brane with $k_i$ strings, 
whereas the D5-brane realization is constructed by identifying $j$-th column of the diagram with $k_j$ boxes 
with $j$-th D5-brane with $k_j$ strings.  

\section{Bubbling geometries}
The supergravity solutions which are dual to the half-BPS Wilson lines appear in the near horizon limit of the brane construction in Type IIB string theory. 
The ten-dimensional geometry is 
\begin{align}
\label{bubbling_geom}
X&=AdS_2\times S^2\times S^4\times \Sigma, 
\end{align}
that is the $AdS_2$ $\times$ $S^2$ $\times$ $S^4$ fibration over a Riemann surface $\Sigma$ 
with boundary $\partial \Sigma$ being the real axis. 
The surface $\Sigma$ can be identified with the lower half-plane in one sheet of a hyperelliptic Riemann surface of genus $g$ as a compactification of the hyperelliptic curve \cite{DHoker:2007mci}
\begin{align}
s^2&=(u-e_1)\prod_{i=1}^g (u-e_{2i}) (u-e_{2i+1}), 
\end{align} 
where $e_{2g+2}=-\infty$ $<$ $e_{2g+1}$ $<$ $\cdots$ $<$ $e_1$ are real branch points. 
The solutions are determined by two real harmonic functions $h_1$ and $h_2$ on $\Sigma$. 
They are non-singular except for a point $u_0>e_1$ on $\partial \Sigma$ corresponding to the asymptotic $AdS_5\times S^5$ region. 
While $h_2$ obeys the Dirichlet boundary condition at $\partial \Sigma$, 
the boundary conditions of $h_1$ change at the $(2g+2)$ points on $\partial \Sigma$ corresponding to the branch points. 
It satisfies the Dirichlet boundary condition in the slits right side of the points $e_{2i-1}$, $i=1, \cdots, g$ 
where the $S^4$ shrinks to zero and the Neumann boundary condition in those left side of them where the $S^2$ vanishes. 

On the degeneration locus of the $S^4$, 
the geometries develop bubbles of $5$-cycles $\mathcal{C}_5^i$, $i=0,\cdots, g+1$ 
where $e_{2g+3}=-\infty$, $e_0=e_{-1}=u_0$, $e_{-2}=\infty$. 
They are formed by the fibration of the $S^4$ over a segment with one endpoint in the interval $(e_{2i+1},e_{2i})$ and the other in $(e_{2i-1}, e_{2i-2})$. 
Also $7$-cycles $\mathcal{C}_7^{i}$ arises as the warped product $S^2\times \mathcal{C}_5^i$. 
On the degeneration locus of the $S^2$, 
bubbles of $3$-cycles $\mathcal{C}_3^j$ are built as the fibration of the $S^2$ 
over a segment with one endpoint in the interval $(e_{2j+2},e_{2j+1})$ and the other in $(e_{2j}, e_{2j-1})$ 
as well as the $7$-cycles $\tilde{\mathcal{C}}_7^{j}$ as the warped product $S^4\times \mathcal{C}_3^j$ with $j=1,\cdots, g$. 

\begin{figure}
\begin{center}
\includegraphics[width=16cm]{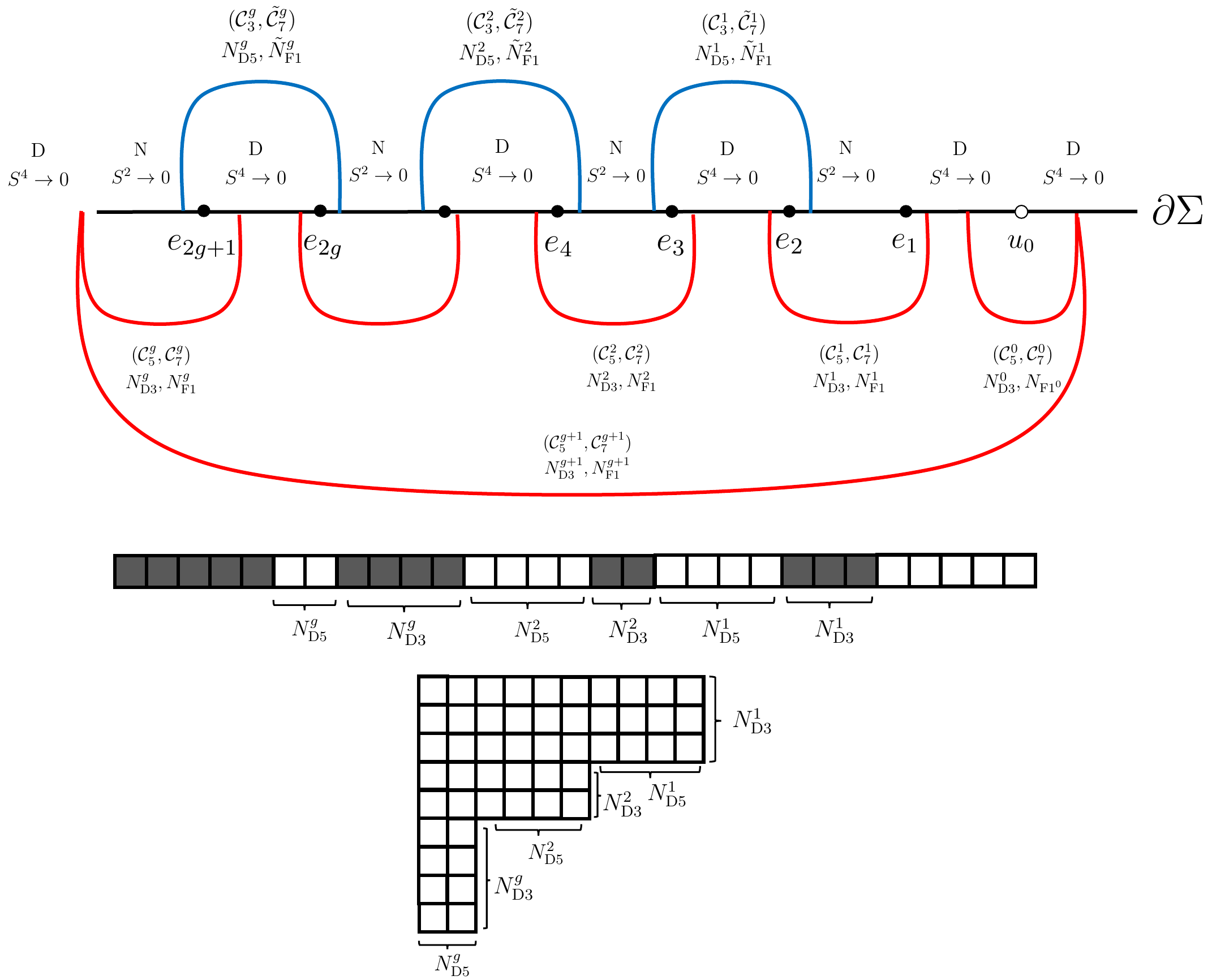}
\caption{The bubbles of cycles appearing along the boundary $\partial \Sigma$ of the Riemann surface $\Sigma$ due to the alternating boundary conditions for $h_1$ 
where N (resp. D) stands for the Neumann (resp. Dirichlet) boundary condition, at which the $S^2$ (resp. $S^4$) shrinks (top). 
The Maya diagram consisting of black boxes corresponding to D3-branes and white boxes to D5-branes (middle). 
The Young diagram associated to the representation for the dual Wilson line (bottom). }
\label{fig_bubbling}
\end{center}
\end{figure}

The D5-brane charges, the D3-brane charges and the fundamental string charges can be computed 
from the supergravity solutions for 
the $3$-cycles $\mathcal{C}_3^j$, $5$-cycles $\mathcal{C}_5^{i}$, and $7$-cycles $\mathcal{C}_7^i$, $\tilde{\mathcal{C}}_7^j$ 
which support the RR $3$-form, the RR $5$-form and the NSNS $3$-form respectively. 
It follows from the explicit computation of the charges in the canonical gauge that 
\cite{Benichou:2011aa} 
\begin{align}
\label{charge_cond}
N_{\textrm{F1}}^{i}&=
N_{\textrm{D3}}^{i}\sum_{j=i}^{g}N_{\textrm{D5}}^{j}, 
\qquad \textrm{for $i=1,\cdots, g$}
\nonumber\\
N_{\textrm{F1}}^0&=\sum_{i=1}^{g}N_{\textrm{D3}}^{i}\sum_{j=i}^{g}N_{\textrm{D5}}^{j}, 
\qquad 
N_{\textrm{F1}}^{g+1}=0,
\nonumber\\
\qquad
N_{\textrm{D3}}^{0}&=N, 
\qquad 
N_{\textrm{D3}}^{g+1}=N-\sum_{i=1}^{g}N_{\textrm{D3}}^i, 
\end{align}
where $N_{\textrm{F1}}^{i}$, $N_{\textrm{D3}}^{i}$ and $N_{\textrm{D5}}^{j}$  
are the number of fundamental strings for the $\mathcal{C}_7^i$, 
that of D3-branes for the $\mathcal{C}_5^i$ 
and that of D5-branes for the $\mathcal{C}_3^j$. 
The conditions (\ref{charge_cond}) verify the identification \cite{Okuda:2008px} of the genus $g$ supergravity solutions (\ref{bubbling_geom})
with the Young diagrams containing $g$ parts in such a way that 
the $\partial \Sigma$ is obtained from the Maya diagram which is one-to-one correspondence with the Young diagram (see e.g. \cite{MR977036}). 
The corresponding Maya diagram contains $N_{\textrm{D3}}^i$ consecutive black cells as the segments for the $i$-th stack of D3-branes 
and $N_{\textrm{D5}}^j$ consecutive white cells for the $j$-th stack of D5-branes except for those far to the left and right. 
The lengths of the vertical (resp. horizontal) segments $|$ (resp. $-$) on the boundary of the Young diagram 
are given by the numbers of D3-branes (reps. of D5-branes) on the corresponding slits on $\partial \Sigma$ (see Figure \ref{fig_bubbling}). 
%
%
%
%
%

\section{Quantum fluctuations}
Our starting point to obtain the spectra of quantum fluctuations of the bubbling geometries (\ref{bubbling_geom}) 
is the Schur index \cite{Gadde:2011ik,Gadde:2011uv} which is a supersymmetric partition function on $S^1\times S^3$. 
The index is protected in the infrared and may depend on two variables $q$ and $t$ 
which are coupled to the charges of the superconformal algebra. 
It can be viewed as the Taylor series in variable $q^{1/2}$ and the Laurent polynomial in variable $t$. 
In \cite{Hatsuda:2022xdv}, we found the following closed-form expression:
\begin{align}
\label{schur_index}
\mathcal{I}^{U(N)}(t;q)&=
-\sum_{p_1<p_2<\cdots<p_N\in \mathbb{Z}}
q^{-\frac{N^2}{4}}
t^{N^2}
\prod_{i=1}^{N}
\frac{q^{\frac{p_i}{2}}t^{-2p_i}}
{1-q^{p_i-\frac{N}{4}}t^{N}}. 
\end{align}
According to the AdS/CFT correspondence \cite{Maldacena:1997re}, in the large $N$ limit the index (\ref{schur_index}) is shown to be equivalent to the multi-particle gravity index 
\cite{Kinney:2005ej}
\begin{align}
\label{largeN_index}
I^{AdS_5\times S^5}(t;q)
&=
\lim_{N\rightarrow \infty}
\mathcal{I}^{U(N)}(t;q)
=\prod_{n=1}^{\infty}
\frac{(1-q^n)}{(1-q^{\frac{n}{2}}t^{2n}) (1-q^{\frac{n}{2}}t^{-2n})}, 
\end{align}
which encodes the BPS spectrum of the quantum fluctuations 
produced by a gas of free gravitons and their superpartners on $AdS_5\times S^5$. 

In the limit $t\rightarrow 1$, one finds the unflavored index whose coefficients count the number of the $1/8$-BPS local operators in $\mathcal{N}=4$ SYM $U(N)$ theory. 
Furthermore, the enhanced $1/2$-BPS sector is obtained by taking the limit $q\rightarrow 0$ while keeping $\mathfrak{q}:=q^{\frac12}t^2$ finite, 
which we call the half-BPS limit of the index \cite{Hatsuda:2022xdv}. 
In the unflavored limit and the half-BPS limit,  
the multi-particle gravity index (\ref{largeN_index}) reduces to
\begin{align}
I^{AdS_5\times S^5}_{\textrm{$\frac18$BPS}}(q)&=\lim_{
\begin{smallmatrix}
N\rightarrow \infty \\
t\rightarrow 1\\
\end{smallmatrix}
}
\mathcal{I}^{U(N)}(t;q)=\prod_{n=1}^{\infty} \frac{1+q^{\frac{n}{2}}}{1-q^{\frac{n}{2}}}
=\sum_{n\ge0}\overline{p}(n) q^{\frac{n}{2}}, \\
I^{AdS_5\times S^5}_{\textrm{$\frac12$BPS}}(\mathfrak{q})
&=\lim_{\begin{smallmatrix}
N\rightarrow \infty\\
\textrm{$\mathfrak{q}:=q^{\frac12}t^2$ finite}\\
q\rightarrow \infty\\
\end{smallmatrix}}
\mathcal{I}^{U(N)}(t;q)
=\prod_{n=1}^{\infty}
\frac{1}{1-\mathfrak{q}^n}
=\sum_{n\ge} p(n)\mathfrak{q}^n, 
\end{align}
which coincides with the generating function for the number $\overline{p}(n)$ of overpartitions \cite{MR2034322} of $n$ 
and that for the number $p(n)$ of partition of $n$ respectively. 

Provided that $\mathcal{N}=4$ SYM theory is placed on $S^1\times S^3$, 
the BPS Wilson lines can wrap the $S^1$ and sit at points along a great circle in the $S^3$ 
so that the Schur index (\ref{schur_index}) can be generalized to 
correlation functions of the line defects \cite{Gang:2012yr,Cordova:2016uwk}. 
According to the supersymmetric localization, 
the (unnormalized) correlator of the Wilson lines in the representations $\mathcal{R}_j$, $j=1,\cdots, k$ 
can be evaluated from the elliptic matrix integral \cite{Gang:2012yr}
\begin{align}
\label{Schur_line0}
\langle W_{\mathcal{R}_1}\cdots W_{\mathcal{R}_k}\rangle^{U(N)}(t;q)
&=
\frac{(-1)^N q^{-\frac{N^2}{4}}t^{N^2}}
{N!}
\oint 
\prod_{i=1}^{N} 
\frac{d\sigma_i}{2\pi is\sigma_i}
\frac{\theta'(1;q)^N \prod_{i\neq j} \theta(\frac{\sigma_i}{\sigma_j};q)}
{\prod_{i,j}\theta(qt^{-4} \frac{\sigma_i}{\sigma_j};q)}
\prod_{j=1}^{k}\chi_{\mathcal{R}_j}(\sigma), 
\end{align}
where $\theta(x;q)$ $:=$ $\sum_{n\in \mathbb{Z}}(-1)^n x^{n+\frac12}q^{\frac{n^2+n}{2}}$, 
$\theta'(x;q)$ $=$ $\partial_x \theta(x;q)$ 
and $\chi_{\mathcal{R}_j}$ is the character of the representation $\mathcal{R}_j$. 
The additional degrees of freedom due to the insertion of line operators can be obtained from the normalized correlator defined by 
\begin{align}
\label{Schur_line}
\langle \mathcal{W}_{\mathcal{R}_1}\cdots \mathcal{W}_{\mathcal{R}_k}\rangle^{U(N)}(t;q)
&:=\frac{
\langle W_{\mathcal{R}_1}\cdots W_{\mathcal{R}_k}\rangle^{U(N)}(t;q)
}{\mathcal{I}^{U(N)}(t;q)}. 
\end{align}

As a pair of the Wilson lines in the irrep $\mathcal{R}$ at a north pole and its conjugate $\overline{\mathcal{R}}$ at a south pole in the $S^3$ 
can form a straight line in the flat space upon the conformal map to preserve a one-dimensional superconformal symmetry \cite{Cordova:2016uwk}, 
we define the Schur line defect index by their $2$-point function
\begin{align}
\label{line_index}
\mathcal{I}_{\mathcal{R}}^{U(N)}(t;q):=\langle  \mathcal{W}_{\mathcal{R}} \mathcal{W}_{\overline{\mathcal{R}}}\rangle^{U(N)}(t;q). 
\end{align}

The direct calculation of the spectra of the excitations around the gravity dual geometry (\ref{bubbling_geom}) 
for the half-BPS Wilson line in the irrep $\mathcal{R}$ is a non-trivial question. 
Here we seek the single particle gravity index defined as a generating function of the BPS spectrum 
\begin{align}
i^{X}(t;q)&:={\Tr} _{\mathcal{H}} 
(-1)^F q^{\frac{h+j}{2}} t^{2(q_2-q_3)}, 
\end{align}
where the trace is taken over the Hilbert space $\mathcal{H}$ of the BPS states obeying the condition $h=j+q_2+q_3$. 
The generators $F$ $h$, $j$ and $q_i$, $i=1,2,3$ are 
the Fermion number operator, the scaling dimension, the $SO(3)$ spin and the $SO(6)$ Cartan generators. 

Similarly to (\ref{largeN_index}), the multi-particle gravity index 
can be obtained from the Schur line defect index (\ref{line_index}) by taking the large $N$ limit
\begin{align}
\label{line_gravity_index}
I^{X}(t;q)
=\lim_{N\rightarrow\infty} \mathcal{I}_{\mathcal{R}}^{U(N)}(t;q). 
\end{align}
Given the multi-particle gravity index, 
the single particle gravity index can be obtained by taking the plethystic logarithm \cite{MR1601666}
\begin{align}
i^{X}(t;q)&=PL[I^{X}(t;q)]:=\sum_{d\ge 1}\frac{\mu(d)}{d}\log \left[ I^{X}(t^d,q^d) \right], 
\end{align}
where $\mu(k)$ is the M\"{o}bius function. 

To proceed with the calculation, we observe that the charged Wilson line correlators 
characterized by the power sum symmetric functions $p_n(\sigma)$ play a role of a critical platform. 
The large $N$ limit of the $2$-point function of the Wilson line of charge $n$ and that of $-n$ is given by \cite{Hatsuda:2023iwi}
\begin{align}
\label{largeN_2pt_ch}
\langle \mathcal{W}_{n} \mathcal{W}_{-n}\rangle^{U(\infty)}
&=\frac{n(1-q^n)}{(1-q^{\frac{n}{2}}t^{2n})(1-q^{\frac{n}{2}}t^{-2n})}. 
\end{align}
For $n=1$ the Wilson line transforms in the fundamental representation. 
The gravity indices read 
\begin{align}
I^{\textrm{string}}(t;q)&=\frac{1-q}{(1-q^{\frac12}t^2)(1-q^{\frac12}t^{-2})}, \\
\label{single_F1}
i^{\textrm{string}}(t;q)&=-q+q^{\frac12}t^2+q^{\frac12}t^{-2}. 
\end{align}
As argued in \cite{Gang:2012yr}, 
the single particle index (\ref{single_F1}) precisely matches the spectrum of the quantum fluctuations of the gravity dual configuration 
calculated in \cite{Drukker:2000ep} 
where the fundamental string wrapping $AdS_2$ and propagating in $AdS_5\times S^5$. 
The term $-q$ is one of the 8 massive fermions with $(h,j,q_2-q_3)$ $=$ $(3/2,1/2,0)$ 
and the terms $q^{1/2}t^2$, $q^{1/2}t^{-2}$ are two of the $5$ massless scalars 
with $(h,j,q_2-q_3)$ $=$ $(1,0,1)$, $(1,0,-1)$ 
describing the fluctuations of the fundamental string in the $S^5$. 

Our strategy to get more general gravity indices follows from the prescription in \cite{Hatsuda:2023imp}. 
We first use the Jacobi-Trudi identity $s_{\lambda}(\sigma)$ $=$ $\det (h_{\lambda_i+j-i}(\sigma))$, 
where $h_k(\sigma)$ is the complete homogeneous symmetric function 
and the Newton's identity $kh_k(\sigma)$ $=$ $\sum_{i=1}^k h_{k-i}(\sigma)p_i(\sigma)$ 
to express the Schur function $s_{\lambda}(\sigma)$, i.e. the character of the irrep labeled by the Young diagram $\lambda$ 
in terms of the power sum symmetric functions. 
Consequently the multi-particle gravity index can be viewed as the large $N$ correlation function of the charged Wilson lines. 
Furthermore, according to the factorization property \cite{Hatsuda:2023iwi}
\begin{align}
\langle 
\prod_{j=1}^{k} 
(\mathcal{W}_{n_j}\mathcal{W}_{-n_j})^{m_j}
\rangle^{U(\infty)}
&=\prod_{j=1}^{k}m_j! \left(
\langle \mathcal{W}_{n_j}\mathcal{W}_{-n_j} \rangle^{U(\infty)}
\right)^{m_j}, 
\end{align}
it is expressible in terms of the large $N$ charged $2$-point functions (\ref{largeN_2pt_ch}). 

When $k$ grows as $k\sim N$, the dual geometries have the probe brane descriptions  
in terms of a D3-brane (resp. D5-brane) with $k$ units of flux wrapping $AdS_2\times S^2$ (resp. $AdS_2\times S^4$). 
The  multi-particle gravity indices are obtained by taking the large $N$ and large $k$ limit while keeping $N/k$ finite
\begin{align}
&
I^{\textrm{probe D3}}(t;q)=
\lim_{
\begin{smallmatrix}
N\rightarrow \infty\\
k\rightarrow \infty\\
\end{smallmatrix}
}
\mathcal{I}^{U(N)}_{(k)}(t;q), \\
&
I^{\textrm{probe D5}}(t;q)=
\lim_{
\begin{smallmatrix}
N\rightarrow \infty\\
k\rightarrow \infty\\
\end{smallmatrix}
}
\mathcal{I}^{U(N)}_{(1^k)}(t;q). 
\end{align}
One finds \cite{Gang:2012yr,Hatsuda:2023iwi}
\begin{align}
&
I^{\textrm{probe D3}}(t;q)=
I^{\textrm{probe D5}}(t;q)=\prod_{n=1}^{\infty}
\frac{1}{(1-q^{\frac{n}{2}}t^{2n}) (1-q^{\frac{n}{2}}t^{-2n})},\\
\label{single_D3D5}
&
i^{\textrm{probe D3}}(t;q)=
i^{\textrm{probe D5}}(t;q)=
\frac{q^{\frac12}t^2}{1-q^{\frac12}t^2}
+\frac{q^{\frac12}t^{-2}}{1-q^{\frac12}t^{-2}}. 
\end{align}
In fact, the single particle index (\ref{single_D3D5}) encodes quantum fluctuations of the gravity dual configurations 
obtained from the action of a curved probe D-brane with flux \cite{Faraggi:2011bb,Faraggi:2011ge}. 
The BPS spectrum of excitations of the probe D3-brane with flux wrapping $AdS_2\times S^2$ in $AdS_5\times S^5$ is given by an infinite number of fields 
as a Kaluza-Klein (KK) tower of scalars with $(h,j,q_2-q_3)$ $=$ $(l+1,0,-l-1+2i)$ describing the embedding of the D3-brane in the $S^5$ 
and that of fermions with $(h,j,q_2-q_3)$ $=$ $(l+3/2,1/2,-l+2i)$, 
where $l=0,1,2,\cdots$; $i=0,1,\cdots, l+\frac{1+(-1)^F}{2}$ \cite{Faraggi:2011bb}. 
\footnote{We note that the $1/8$- and $1/2$-BPS indices perfectly agree with the spectrum in \cite{Faraggi:2011bb} with the $SO(3)$ quantum numbers $j=0$ for scalars and $j=1/2$ for fermions rather than $j=l$ and $j=l+1/2$.}
Likewise, the BPS spectrum of fluctuations of the probe D5-brane with flux wrapping $AdS_2\times S^4$ in $AdS_5\times S^5$ contains  
the same set of KK towers \cite{Faraggi:2011ge} (see \cite{Gang:2012yr} for the detail). 

When $k$ grows as $k\sim N^2$, the dual geometries are fully backreacted as bubbling geometries. 
So far the calculation of excitations on the bubbling geometries from the gravity side is out of reach due to the invalidity of the probe D-brane approximation. 
Nevertheless, we can still address them from the dual gauge theory side by following the above method! 

\begin{figure}
\begin{center}
\includegraphics[width=7.5cm]{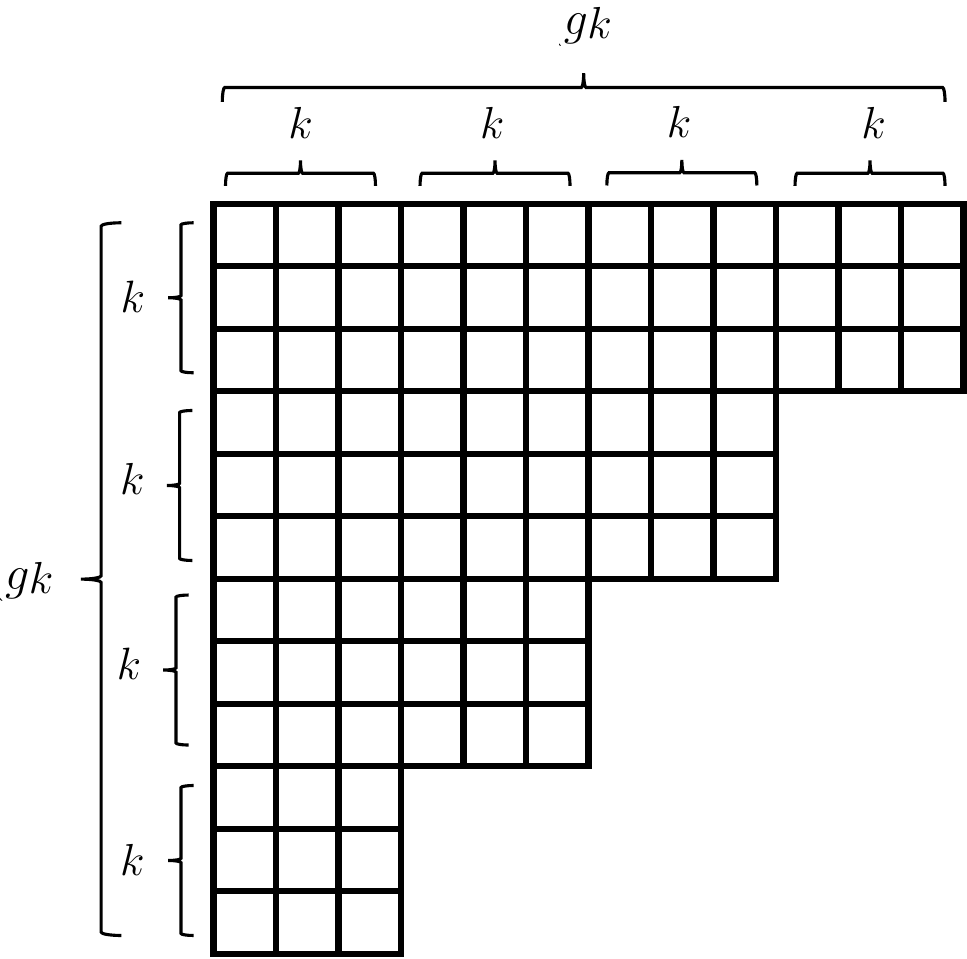}
\caption{The Young diagram $((gk)^k,$ $((g-1)k)^k,$ $\cdots,$ $k^k)$. }
\label{fig_young}
\end{center}
\end{figure}

The quadratic area growth of boxes of the Young diagram for the bubbling geometry of genus $g$ 
can be realized by the Young diagram $((gk)^k,$ $((g-1)k)^k,$ $\cdots,$ $k^k)$ consisting of $(k^k)$'s, the Young diagrams of square shape (see Figure \ref{fig_young}). 
By taking the large $k$ and large $N$ limit of the Schur line defect index for it, we get the following elegant form of the multi-particle gravity index for the bubbling geometry of genus $g$
\begin{align}
\label{largeN_bubbling1}
I_g^{\textrm{bubbling}}(t;q)
&=\lim_{k\rightarrow \infty}
\mathcal{I}_{((gk)^k, ((g-1)k)^k, \cdots, k^k)}^{U(\infty)}(t;q)
\nonumber\\
&=\prod_{n=1}^{\infty}
\frac{(1-q^{\frac{n}{2}} t^{2n}) (1-q^{\frac{n}{2}} t^{-2n})}
{1-(g+1)(t^{2n}+t^{-2n})q^{\frac{n}{2}}+(2g+1)q^{n} }. 
\end{align}

In the unflavored limit $t\rightarrow1$ and the half-BPS limit, 
the expression (\ref{largeN_bubbling1}) reduces to 
\begin{align}
I_g^{\textrm{bubbling}}(q)&=
\prod_{n=1}^{\infty}\frac{1-q^{\frac{n}{2}}}
{1-(2g+1)q^{\frac{n}{2}}}, \\
I_{g,\textrm{$\frac12$BPS}}^{\textrm{bubbling}}(\mathfrak{q})&=
\prod_{n=1}^{\infty}
\frac{1-\mathfrak{q}^n}{1-(g+1)\mathfrak{q}^n}. 
\end{align}
The function
\begin{align}
\prod_{n=1}^{\infty}
\frac{1-q^n}{1-rq^n}=\sum_{n\ge0} C_{n,r}q^n
\end{align}
is the generating function for the number $C_{n,r}$ of conjugacy classes of a general linear group $GL(n,r)$ over a finite field with $r$ elements \cite{MR109810,MR615131}. 
Hence the spectrum of the quantum fluctuations of the bubbling geometry with genus $g$ 
over the multi-particle $1/8$-BPS (resp. $1/2$-BPS) states at level $n$ exactly agrees with $C_{n,2g+1}$ (resp. $C_{n,g+1}$)! 
\footnote{This generalizes the result for $g=1$ in \cite{Hatsuda:2023imp}. }

From (\ref{largeN_bubbling1}) we obtain the single particle gravity index for the bubbling geometry of genus $g$
\begin{align}
\label{gindex_bubbling1}
&i^{\textrm{bubbling}}_g(t;q)
=-\frac{q^{\frac12}t^2}{1-q^{\frac12}t^2}-\frac{q^{\frac12}t^{-2}}{1-q^{\frac12}t^{-2}}
\nonumber\\
&+\sum_{n=1}^{\infty}
\sum_{\begin{smallmatrix}
m_1,m_2,m_3\ge0\\
(m_1,m_2,m_3)\neq (0,0,0)\\
\end{smallmatrix}}
(-1)^{m_3}
N\left( (g+1)^{m_1+m_2} (2g+1)^{m_3},n \right)
M(m_1,m_2,m_3)
\nonumber\\
&\times 
q^{\left( \frac{m_1+m_2}{2}+m_3 \right)n}
t^{2(m_1-m_2)n}. 
\end{align}
Here 
\begin{align}
N(m,n)&=\frac{1}{n}\sum_{d|n}\varphi(d)m^{\frac{n}{d}}
\end{align}
is the general necklace polynomial, 
the number of necklaces of length $n$ consisting of beads with $m$ distinct colors (see e.g. \cite{MR1676282})  
where $\sum_{d|n}$ is the sum over divisors $d$ of $n$ and $\varphi(n)$ is the Euler's totient function. 
Also 
\begin{align}
M(m_1,m_2,m_3)&=
\frac{1}{m_1+m_2+m_3}
\sum_{m|\mathrm{gcd}(m_1,m_2,m_3)}
\mu(m)
\left(
\begin{matrix}
\frac{m_1+m_2+m_3}{m}\\
\frac{m_1}{m},\frac{m_2}{m},\frac{m_3}{m}\\
\end{matrix}
\right)
\end{align}
is the number of circular words of length $(m_1+m_2+m_3)$ 
and minimal period $(m_1+m_2+m_3)$ with letter $x_i$ appearing $m_i$ times \cite{MR2143453}. 

In the unflavored limit $t\rightarrow1$, 
the gravity index (\ref{gindex_bubbling1}) becomes
\begin{align}
\label{gindex_bubbling2}
i^{\textrm{bubbling}}_g(q)
&=
\sum_{n=1}^{\infty}\Bigl[N(2g+1,n)-1\Bigr]q^{\frac{n}{2}}. 
\end{align}
Similarly, the half-BPS limit of the gravity index (\ref{gindex_bubbling1}) reads
\begin{align}
\label{gindex_bubbling3}
i^{\textrm{bubbling}}_{g,\textrm{$\frac12$BPS}}(\mathfrak{q})
&=\sum_{n=1}^{\infty} \Bigl[N(g+1,n)-1\Bigr]\mathfrak{q}^n. 
\end{align}
The expression (\ref{gindex_bubbling2}) (resp. (\ref{gindex_bubbling3})) demonstrates that 
the degeneracy of the excitations for the $1/8$-BPS (resp. $1/2$-BPS) single particle states on the bubbling geometry of genus $g$ 
is equal to the number of necklaces of length $n$ whose beads with $(2g+1)$ (resp. $(g+1)$) distinct colors minus one! 

It is worth mentioning that the bubbling geometry exhibits a new class of asymptotic degeneracy of states. 
When the multi-particle gravity indices are expanded as
\begin{align}
I^{X}(q)&=\sum_{n\ge0}d^{X}(n)q^{\frac{n}{2}},\\
I^{X}_{\textrm{$\frac12$BPS}}(\mathfrak{q})&=\sum_{n\ge0}d_{\textrm{$\frac12$BPS}}^{X}(n) \mathfrak{q}^n, 
\end{align}
the coefficient $d^{X}(n)$ (resp. $d_{\textrm{$\frac12$BPS}}^{X}(n)$) is the degeneracy of the $1/8$-BPS (resp. $1/2$-BPS) multi-particle states at level $n$ for the geometry $X$. 
As $n\rightarrow \infty$, the degeneracies behave as
\begin{align}
d^{\textrm{string}}(n)&\sim 2,
&d_{\textrm{$\frac12$BPS}}^{\textrm{string}}(n)\sim 1, \\
\label{asymp_brane}
d^{\textrm{probe D$p$}}(n)&\sim 
\frac{1}{4\cdot 3^{\frac34} n^{\frac54}} \exp\left[ \frac{2\pi}{3^{\frac12}} n^{\frac12} \right], 
&d_{\textrm{$\frac12$BPS}}^{\textrm{probe D$p$}}(n)\sim 
\frac{1}{4\cdot 3^{\frac12} n} \exp\left[ \frac{2^{\frac12}\pi}{3^{\frac12}} n^{\frac12} \right], \\
\label{asymp_bubbling}
d^{\textrm{bubbling}}_{g}(n)&\sim 
\exp\left[\log(2g+1) n\right],
&d_{\textrm{$g, \frac12$BPS}}^{\textrm{bubbling}}(n)\sim 
\exp\left[\log(g+1) n\right],
\end{align}
with $p=3,5$. 
The asymptotic degeneracies (\ref{asymp_bubbling}) for bubbling geometries take the form $\sim \exp(\alpha n)$ where $\alpha$ is some real constant. 
Such an enormous leading behavior is reached as the maximal case 
from the degeneracy for $d$-dimensional free scalar field theory \cite{Cardy:1991kr} in the limit $d\rightarrow\infty$ 
or that for fluctuations of a $p$-brane \cite{Fubini:1972mf, Dethlefsen:1974dr,Strumia:1975rd,Alvarez:1991qs,Harms:1992jt} in the limit $p\rightarrow \infty$. 
Such infinite dimensional disasters indicate that the quantum fluctuations of bubbling geometries need some new class of description. 
Also it would be interesting to explore the different limits which interpolate between the probe brane limit (\ref{asymp_brane}) and the bubbling geometry (\ref{asymp_bubbling}). 

\subsection*{Acknowledgements}
The authors are grateful to Hai Lin for useful discussions and comments. 
The work of Y.H. was supported in part by JSPS KAKENHI Grant Nos. 22K03641 and 23H01093. 
The work of T.O. was supported by the Startup Funding no. 4007012317 of the Southeast University. 

\bibliographystyle{utphys}
\bibliography{ref}

\end{document}